\documentclass[%
 reprint,
 amsmath,amssymb,
 aps,
]{revtex4-1}

\usepackage{dblfloatfix}
\usepackage{wrapfig}
\usepackage[utf8]{inputenc}
\usepackage{graphicx}
\usepackage{dcolumn}
\usepackage{bm}
\usepackage{ulem}
\usepackage{xcolor}
\usepackage{lipsum}
\usepackage{hyperref}
\hypersetup{
    colorlinks=true,
    linkcolor=blue,
    filecolor=magenta,      
    urlcolor=black,
    citecolor=blue,
    pdftitle={Overleaf Example},
    pdfpagemode=FullScreen,
    }

\begin{document}

\preprint{APS/123-QED}

\title{High-resolution broadband characterization of resonance dispersion in an optical microresonator}

\author{Romain Dalidet$^{1}$, Adrien Bensemhoun$^{1}$, Gregory Sauder$^{1}$, Anthony Martin$^{1}$, David Medina$^{2}$, Carlos Alonso-Ramos$^{2}$, Éric Cassan$^{2}$, Laurent Vivien$^{2}$, Jonathan Faugier-Tovar$^{3}$, Baptiste Routier$^{3,4}$, Quentin Wilmart$^{3}$, Ségolène Olivier$^{3}$, Virginia D'Auria$^{1}$, Laurent Labonté$^{1}$}
\email{laurent.labonte@univ-cotedazur.fr}
\author{Sébastien Tanzilli$^{1}$}
\affiliation{Université Côte d’Azur, CNRS, Institut de Physique de Nice, 17 rue Julien Lauprêtre, 06200 Nice, France}
\affiliation{Centre de Nanosciences et de Nanotechnologies, CNRS, Université Paris-Saclay, Palaiseau 91120, France}
\affiliation{Université de Grenoble Alpes, CEA, LETI, F38000 Grenoble, France}
\affiliation{Institut des Nanotechnologies de Lyon, CNRS-UMR-5270 Lyon, Ecole Centrale de Lyon, Ecully, France}

\begin{abstract}
Accurate knowledge of the uneven free spectral range of an optical microresonator, which provides direct insight into group velocity dispersion, is essential for understanding and controlling Kerr frequency comb dynamics. In this work, we present a simple and highly precise method for measuring the free spectral range over a 5~THz bandwidth in silicon nitride microresonators, leveraging a wavemeter with 0.4 MHz resolution. Our fully fibered plug-and-play experimental setup enables the accurate extraction of resonance frequencies. By carefully analyzing the spectral position of each resonance, we measure both second- and third-order free spectral range expansion coefficients. This approach offers a robust and accessible tool for dispersion characterization in integrated photonic circuits, paving the way for next-generation of Kerr comb sources and quantum photonic technologies.

\end{abstract}

\maketitle

Kerr frequency combs generated in optical microresonators (MRs) with small mode volumes and high Q-factors find applications in spectroscopy~\cite{suh_microresonator_2016, yang_vernier_2019}, optical clocks~\cite{Papp:14,Newman:19}, coherent communications~\cite{pfeifle_coherent_2014, yu_silicon-chip-based_2018}, and quantum optics~\cite{kues_quantum_2019, labonte_integrated_2024, bensemhoun_multipartite_2025}. In this context, group velocity dispersion (GVD) which manifests as deviations from an equidistant free spectral range (FSR), is a critical parameter for achieving broad spectral widths and soliton comb operation~\cite{herr_universal_2012, chembo_kerr_2016, liu_monolithic_2020, del_haye_optical_2007, karpov_universal_2017, chang_integrated_2022}. Material and geometric dispersion can be engineered through coatings and resonator geometries~\cite{fujii_dispersion_2020, zhang_dispersion_2021}. Tailoring the resonator geometry can overcome material dispersion, achieve anomalous dispersion for soliton formation, and explore higher-order dispersion for spectral broadening~\cite{herr_temporal_2014}. Additionally, phenomena such as Cherenkov radiation and clustered comb formation further emphasize the need for precise dispersion knowledge~\cite{elgin_perturbative_1995, akhmediev_cherenkov_1995, austin_dispersive_2006, arteaga-sierra_multi-peak-spectra_2014, dudley:hal-00476072, DW_11}.

Common approaches for dispersion measurement include electro-optic sideband spectroscopy ~\cite{Li:12} or frequency-comb-based spectroscopy, which creates an optical ruler through heterodyning~\cite{delhaye_frequency_2009, Fujii:19, Twayana:21, Liu:18}. While these approaches offer high precision by transferring the stability of optical frequency combs, they remain costly and cumbersome, and require numerous accessible resonances, typically achieved in whispering gallery mode resonators with FSRs of a few tens of MHz~\cite{fujii_dispersion_2020, zhang_dispersion_2021, Nielsen18}. For devices with FSR above 100 GHz, the number of accessible resonances significantly decreases, imposing stricter precision requirements. In this context, integrated silicon and silicon nitride MRs offer key advantages, notably FSR values that align with the International Telecommunication Union (ITU) grid, making them highly relevant for both classical~\cite{bernabe_silicon_2021} and quantum communications~\cite{oser_high-quality_2020, PhysRevApplied.12.034053} as well as for frequency comb generation~\cite{herr_universal_2012, bensemhoun_multipartite_2025}. Despite these advantages, existing dispersion measurement techniques generally provide access to the GVD at a single wavelength, usually centered within the analysis band.

Here, we present a direct experimental measurement of the FSR in a silicon nitride MR over a 5~THz bandwidth within the C- and L-telecom bands. Our approach builds upon a standard method based on the Taylor expansion around the central wavelength of the analysis band. Through meticulous spectral analysis, incorporating both single and, more importantly, double resonances, we precisely determine the central frequency of each resonance. This allows for precise extraction of the coefficient of the FSR at multiple measurement points. This extends the scope of conventional methods and sets our work apart from most previous studies, which typically provide only a single measurement point~\cite{Twayana:21, Liu:18}.

The MR dispersion is derived from the Taylor expansion of the resonance frequencies, whose relative positions are determined as

\begin{equation}\label{eq: dispersion microring}
    \omega_\mu = \omega_0 + \sum_{m=0}^{\infty} \frac{1}{m!}D_m \mu^m = \omega_0 + D_1\mu + D_{int}(\mu) \, .
\end{equation}
Here, $\omega_0$ is the central frequency and $\mu \in \mathbb{Z}$ is the relative resonance number. Coefficient $D_1$ characterizes the equidistant FSR of the MR, as it would be without any dispersion. Higher order coefficients, regrouped and summed under the coefficient $D_{int}$, characterize the FSR of a realistic MR under the influence of the dispersion. By carefully measuring the experimental positions of the MR resonances, $D_{int}$ can be extracted to infer the dispersion of the resonator. In particular, the second-order dispersion coefficient $D_2$ is directly related to GVD:

\begin{equation}\label{eq: D2 microring}
    \beta^{(2)} = \frac{-n_{eff}(\omega_0)}{c}\frac{D_2}{D_1^2} \, ,
\end{equation}
where $\beta^{(2)}=\left.\frac{\partial^2 k}{\partial \omega^2}\right|_{\omega_0}$ stands as the GVD and $n(\omega_0)$ is the effective  index of the mode propagating in the waveguide. The last equation is also directly related to the chromatic dispersion:

\begin{equation}\label{eq: D2omega microring}
D_{2\omega} = \frac{-\omega_0^2}{2\pi c}\beta^{(2)} \, ,
\end{equation}
usually expressed in $ps/(nm.km)$. Similarly, the third-order dispersion and dispersion slope \textit{i.e.} the derivative of $D_{2\omega}$, is expressed as

\begin{align}
    \beta^{(3)} & =\frac{3c}{n(\omega)}\beta^{(2)} - \frac{n(\omega_0)}{c}\frac{D_3}{D_1^3} \approx \frac{-n(\omega_0)}{c}\frac{D_3}{D_1^3} \, , \label{eq: D3 microring} \\
    D_{3\omega} & = \Bigl(\frac{\omega_0^2}{2\pi c}\Bigr)^2\beta^{(3)} \, . \label{eq: D3omega microring}
\end{align}

It should be noted that although every high-order dispersion coefficient contributes to the deviation from an equidistant FSR, the relation $D_{n} \gg D_{n+1}$ generally holds for a wide range of optical waveguides. This assumption restricts our analysis to third-order dispersions ~\cite{fujii_dispersion_2020}.

\begin{figure}[h]
    \centering
    \includegraphics[width=1\linewidth]{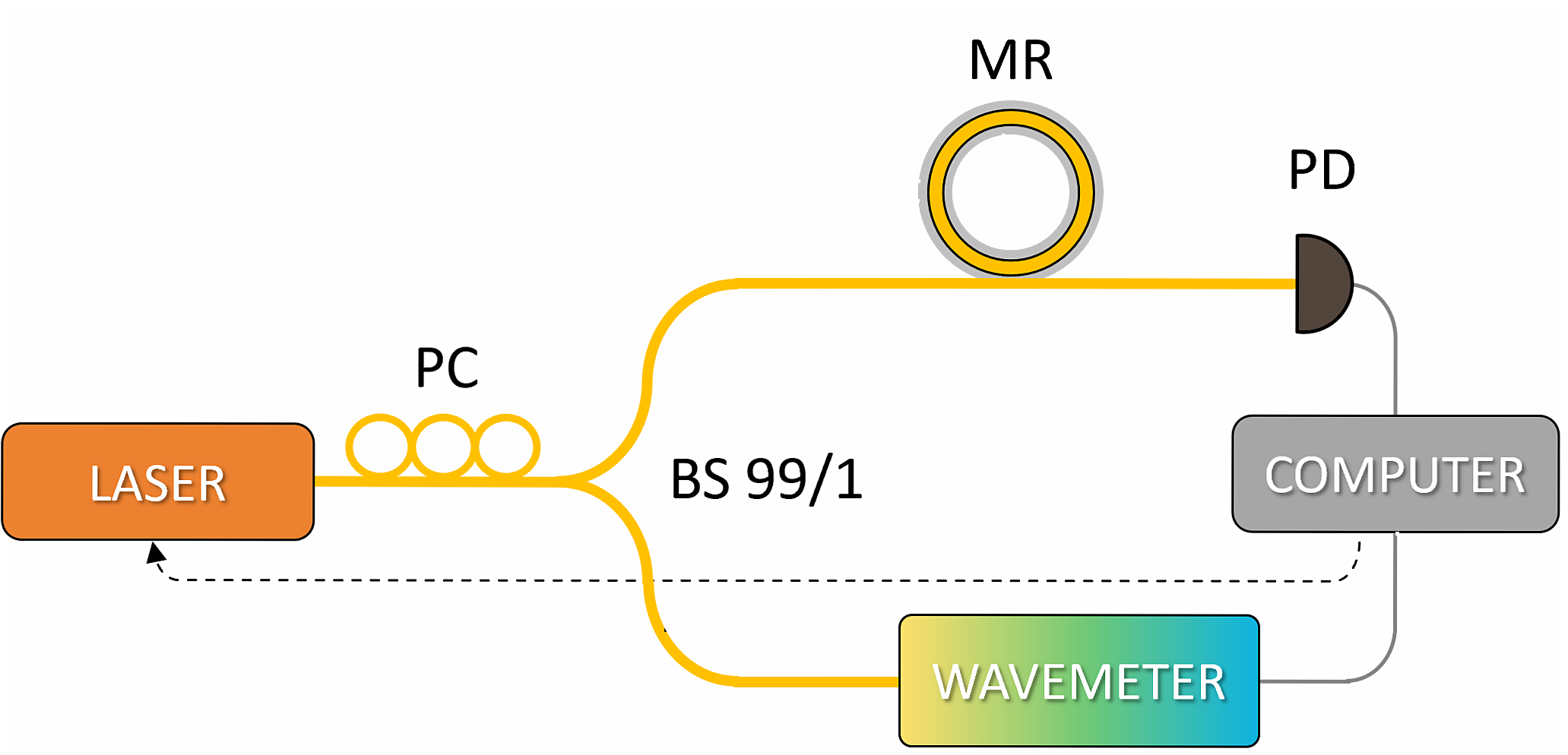}
    \caption{Experimental setup for dispersion characterization of the resonator. To determine the approximate position of the resonances, the laser and the diode are replaced by a superluminescent diode and an optical spectrum analyzer, respectively. PC: polarization controller, MR: microresonator, PD: photodiode, BS: beam-splitter}
    \label{fig: experimental setup}
\end{figure}

The schematic of the experimental setup for measuring the MR dispersion is illustrated in Fig. \ref{fig: experimental setup}. A single-line continuous tunable telecom laser ($\Delta \nu < 10\, KHz$) is used to probe the system. The light beam is split into two paths by using an unbalanced (99/1) fiber beam splitter.

The lower-power output is routed to a wavemeter used as a reference for the measurement (WS8-10, HighFinesse). The wavemeter is based on a Fizeau interferometer and has an accuracy and resolution of 8 MHz and 0.4 MHz, respectively. As the probe laser scan is continuous, the wavelength measurement precision is determined by the resolution of the wavemeter, i.e. 0.4 MHz. The light at the second output of the beam splitter is coupled to the MR and collected at its output using microlensed fibers. The MR is made of silicon nitride with a radius $R \approx 200\,\mu m$ corresponding to an FSR of  $\approx120\,GHz$ and is mounted on a thermo-electric cooler to stabilize its temperature.

The polarization at the resonator input is carefully adjusted using a fiber polarization controller to excite either the transverse electric (TE) or transverse magnetic (TM) mode of the MR. Because the TE and TM modes exhibit different spatial geometries mainly induced by the cross section of the MR, we expect variations in their dispersion measurements. The output power of the MR is monitored using a standard calibrated photodiode.

\begin{figure}[h]
    \centering
    \includegraphics[width=0.9\linewidth]{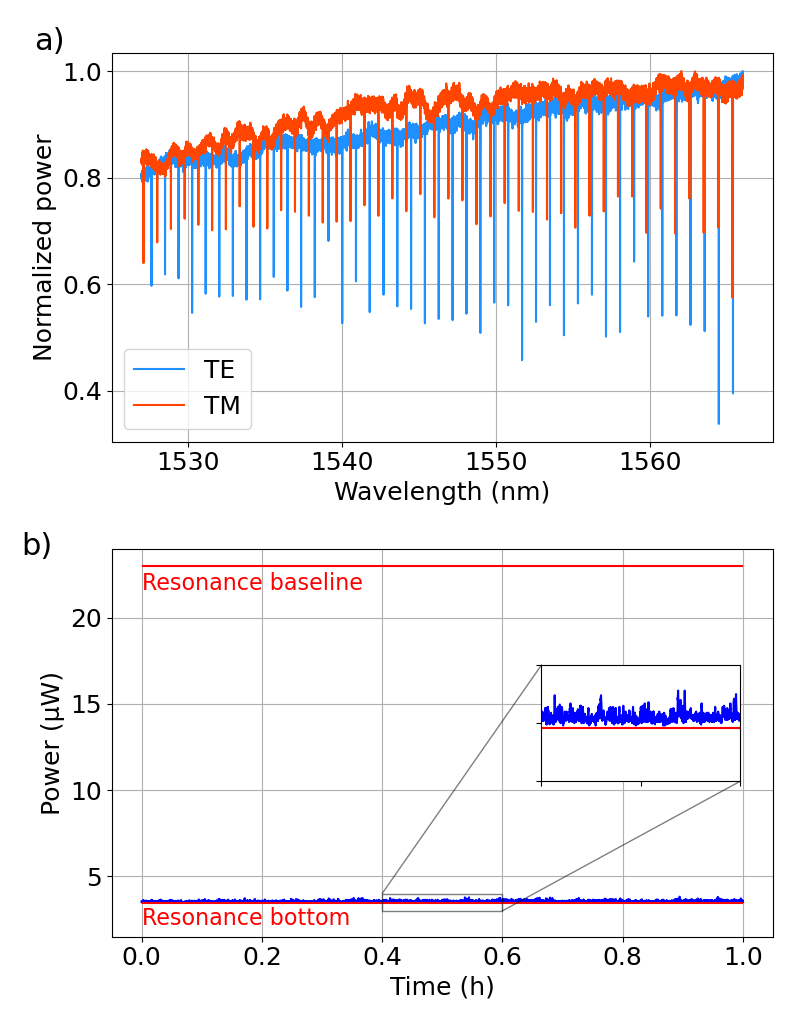}
    \caption{a) TE (blue) and TM (orange) transmission spectrum of the MR using a superluminescent diode. b) Long term stability (blue line) when the pump laser frequency is set at a bottom of an arbitrary resonance.}
    \label{fig: stability}
\end{figure}

The main challenge lies in measuring the positions of narrow spectral lines over a broad spectral range with the highest accuracy. To reduce the measurement time while leveraging the stability of the setup, we proceed in two steps. The first step consists of a pre-alignment procedure to roughly identify the position of each resonance by replacing the laser with a superluminescent diode in the telecom range and the photodiode with an optical spectrum analyzer. This step allows us to determine the approximate resonance positions $\omega_k$ for both TE and TM modes. The measured spectra are shown in Figure \ref{fig: stability}a. In the second step, the computer controlling the experiment drives the probe laser, tuning its frequency to the previously determined $\omega_k$ and subsequently scanning each resonance in a synchronized manner between the photodiode and wavemeter. The frequency tunability of the probe laser enables the measurement of 42 resonances. As each resonance requires approximately 3 s to be measured due to the laser scanning time, the total measurement duration for a given transverse mode is approximately 2 minutes. To assess the stability of the setup and validate the experimental protocol over this timescale, we performed long-duration acquisition, as shown in Fig. \ref{fig: stability}b. The probe laser frequency was set to the minimum of an arbitrary resonance, and the system was allowed to evolve freely. The small amplitude of power variations relative to the baseline of the selected resonance over the 1-hour measurement, whose standard deviation is below 0.2~\%, confirms the stability of both the probe laser and MR.

A significant challenge in accurately determining GVD arises from the presence of double peaks in certain resonances~\cite{janz_measurement_2024}. The bifurcation of the resonance peak is attributed to the backward-propagating waves generated by light scattering in the coupling region or surface roughness~\cite{Matres:17}. Moreover, the two resonance peaks may exhibit asymmetry in both width and amplitude. To account for both potential double peaks and asymmetry in each measured resonance, we use a double asymmetric Lorentzian function as the fitting model:

\begin{align}
    f(\omega)  & = \sum_i\Bigl[y_i - A_i\frac{\gamma_i(\omega)}{\frac{1}{4}\gamma_i(\omega)^2 + (\omega-\omega_i)^2}\Bigr]\, , \label{eq: resonance fit}\\ 
    \gamma_i(\omega) & = \frac{2\gamma_i}{1+e^{\alpha_i(\omega-\omega_i)}} \, . \label{eq: resonance fit gamma}
\end{align}

\begin{figure}[h]
    \centering
    \includegraphics[width=0.95\linewidth]{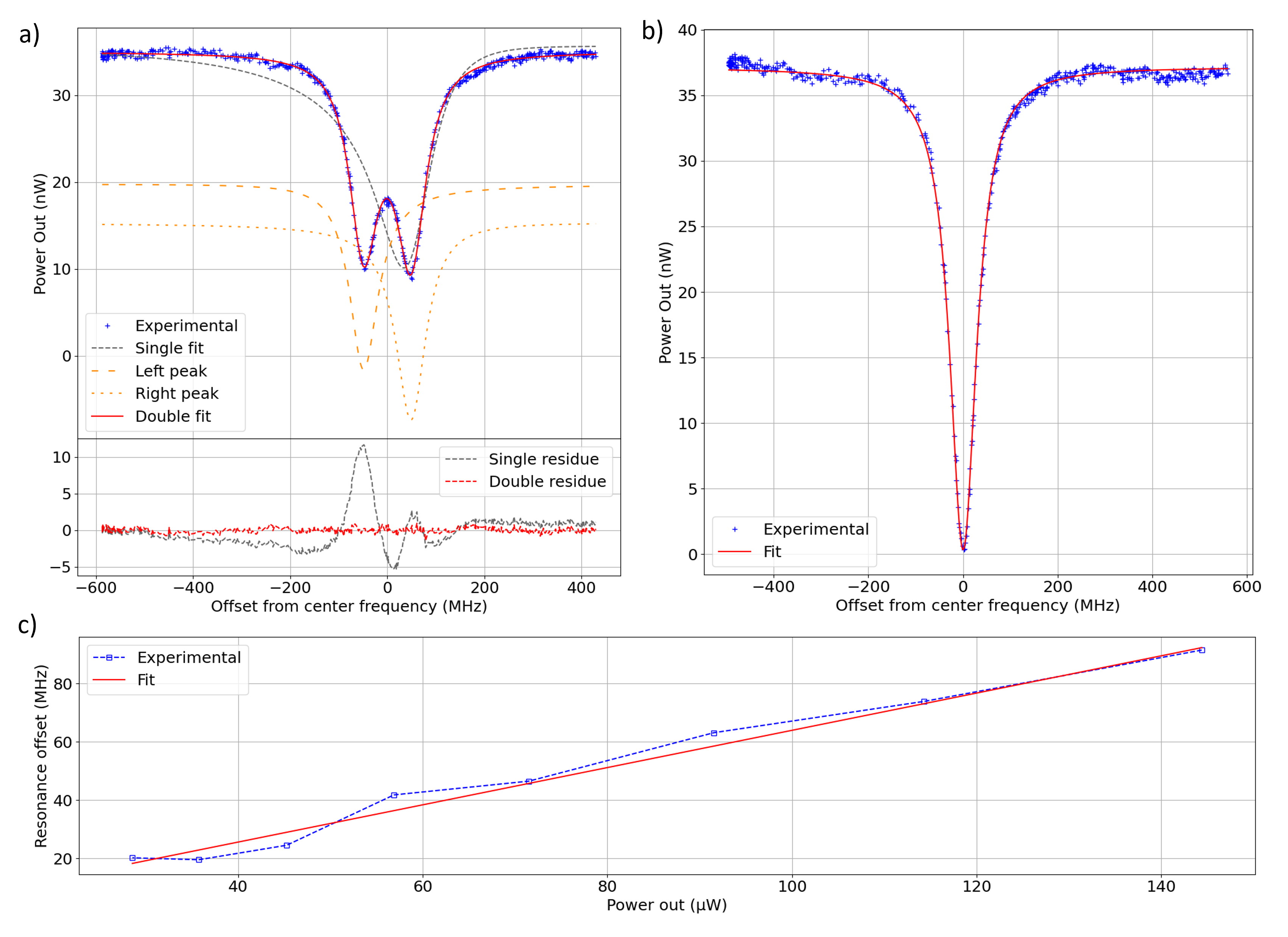}
    \caption{a) Measured double peak resonance as a function of the frequency. The top figure shows the experimental data (blue curve) and associated fit (red curve) using Equation \ref{eq: resonance fit}. The grey curve shows a fit using the simple resonance model, whereas the orange curves show the two compounds of the double resonance model. The bottom figure shows the residuals when one or two peaks are considered in the fit in gray and red, respectively. b) Measured single-peak resonance and fit associated using equation \ref{eq: resonance fit}. c) Resonance center (redshift) as a function of laser probe power. The linear dependence is mainly induced by Kerr and thermo-refractive effects.}
    \label{fig: measured résonances}
\end{figure}

In Eq. \ref{eq: resonance fit}, the index i can take the values "r" or "l for right and left, respectively. Parameter $y_i$ represents the off-resonance transmission (baseline) of the chip, $A_i$ is the amplitude of the Lorentzian function, and $\omega_i$ is its center. In Eq. \ref{eq: resonance fit gamma}, $\alpha_i$ denotes the asymmetry coefficient, while $\gamma_i$ corresponds to the full width at half maximum (FWHM) of the function. Notably, for single resonance, the sum of Eq. \ref{eq: resonance fit} vanishes (the index i takes the value "s" for single), and the asymmetry parameter reduces to $\alpha_s = 0$, yielding the standard Lorentzian function. An example of measured and fitted double-peak and single-peak resonance is shown in Fig. \ref{fig: measured résonances}a and Fig. \ref{fig: measured résonances}b, respectively.
\mbox{}

\onecolumngrid

\newpage
\twocolumngrid
\noindent According to this formalism, the position of $\omega_s$ is directly extracted for a single-peak resonance. In the case of double-peak resonance, where the peaks do not have the same amplitude and FWHM, $\omega_s$ is taken as the mean value of $\omega_l$ and $\omega_r$~\cite{janz_measurement_2024}. \\

Another form of asymmetry can typically arise from an excessively high input laser power, affecting both single- and double-peak resonances. Indeed, an elevated input power induces a red shift in the resonance primarily because of the combined effects of the Kerr effect, thermo-refractive effect, and free carrier absorption~\cite{Luo:12, Wang:13}. 
\begin{figure}[!h]
    \centering
    \includegraphics[width=0.9\linewidth]{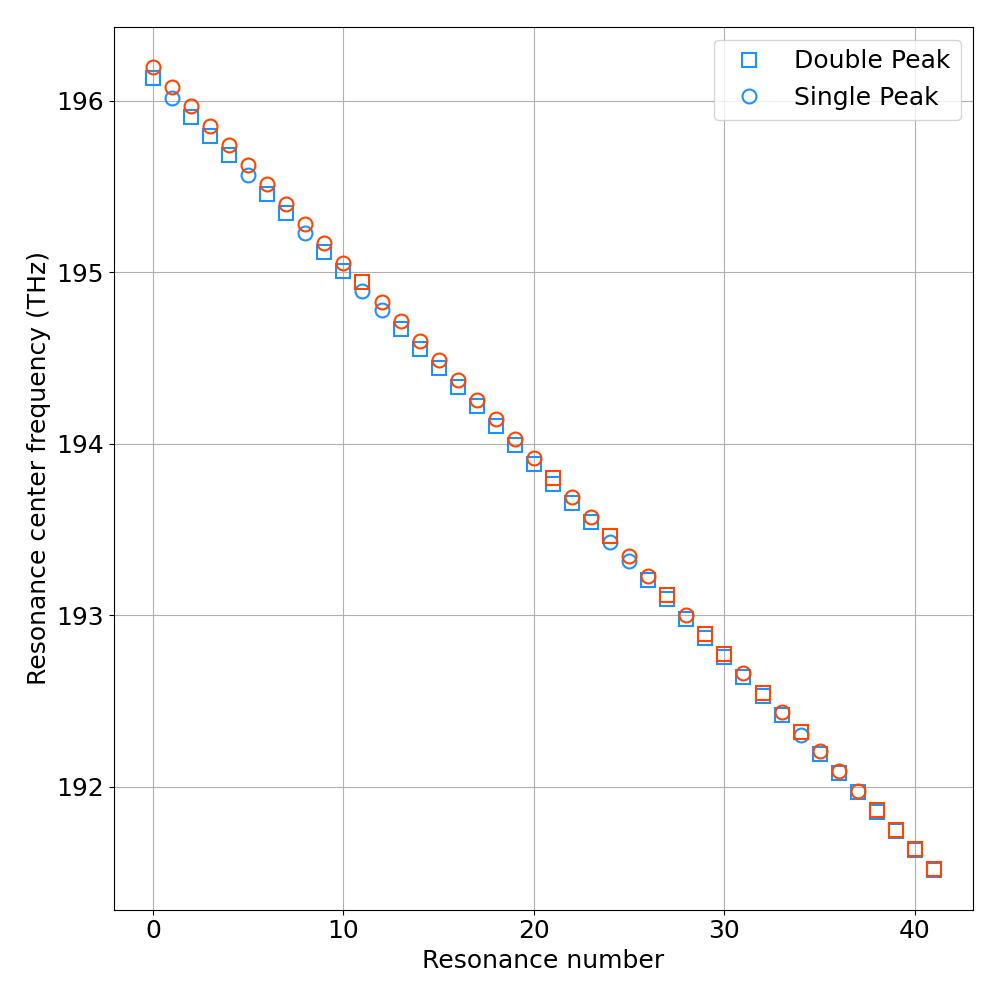}
    \caption{Extracted resonance centers using equation \ref{eq: resonance fit} for TE (blue) and TM (orange) modes. The circle (square) markers represent a simple (double) peak resonance.}
    \label{fig: res résonances}
\end{figure}
As the center of an optically power-induced resonance shift deviates from its original position, an additional error is introduced into the FSR measurement, which consequently affects the accuracy of the GVD determination. Avoiding both the aforementioned asymmetry and resonance center displacement requires two key considerations. First, the rise time of the photodetector must be shorter than the duration required to sweep across resonance. This condition is fulfilled in our setup because each resonance scan takes several seconds, preventing distortion or artefacts that could arise from a mismatch between the scanning speed and data acquisition. Second, the intracavity power must be kept sufficiently low to prevent any redshift of the resonance, ensuring that nonlinear effects such as the Kerr effect, thermo-refractive shifts, and free carrier absorption remain negligible. \\
To determine the intracavity power threshold at which the redshift of the resonances becomes negligible relative to the FSR, we experimentally extracted the center frequency of an arbitrary resonance as a function of the measured probe laser power at the MR output. The results are shown in Fig. \ref{fig: measured résonances} c. The data were acquired for input probe laser powers below the nonlinear threshold of the MR, where the redshift is expected to exhibit linear dependence \cite{shetewy_demonstration_2024}. Furthermore, at a very low input power, the intra-cavity energy remains insufficient to induce any measurable shift in resonance. From the extracted linear coefficient, we empirically set the measured maximum output pump power of the chip outside of the resonance to $<50 \, nW$, inducing a theoretical redshift far below the resolution of the wavemeter.\\

\begin{figure}[!h]
    \centering
    \includegraphics[width=1\linewidth]{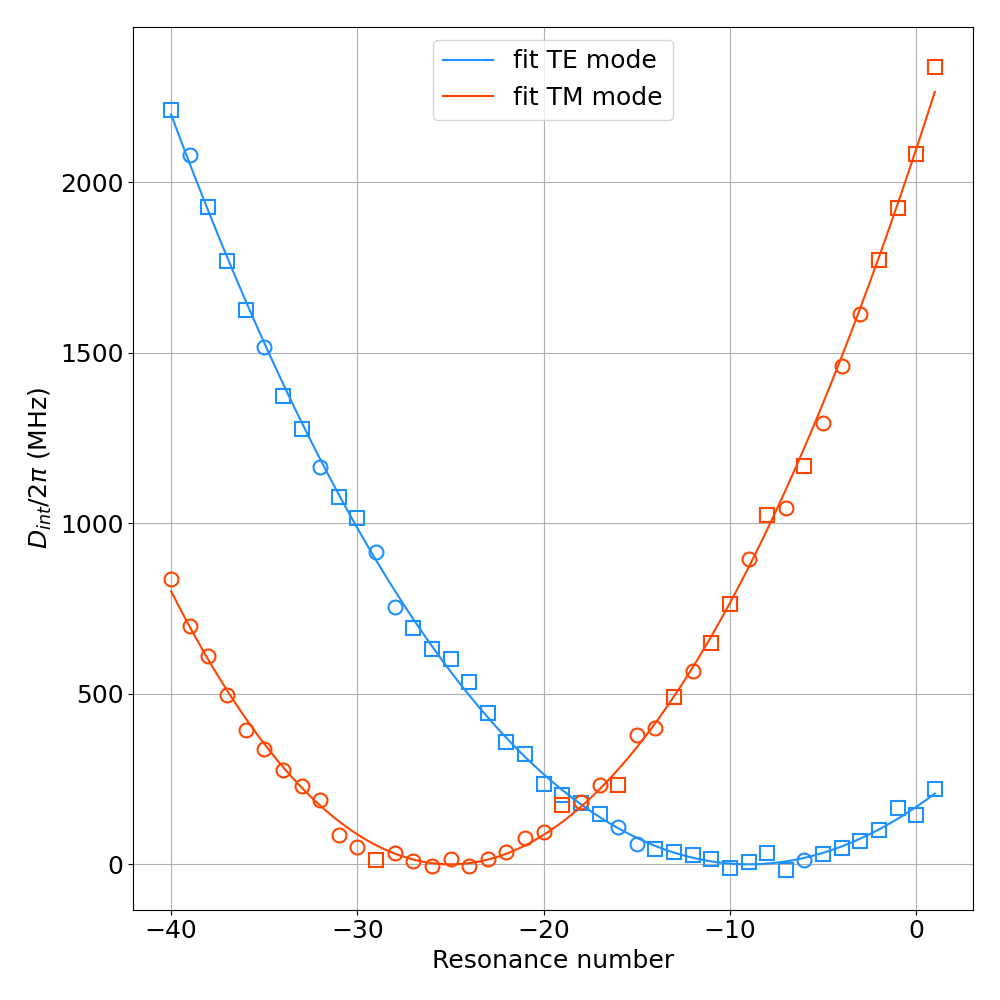}
    \caption{Experimental and fitted $D_{int}$ using equation \ref{eq: dispersion microring} for two arbitrary resonances in TE mode (blue) and TM mode (orange). The linear part, $D_1$ of the equation was subtracted. The circle (square) markers represent a simple (double) peak resonance.}
    \label{fig: dint}
\end{figure}
Each resonance is scanned and fitted according to its single or double peak behavior to extract its center position (see Fig. \ref{fig: res résonances}). We emphasize that the mean FWHM measured across all resonances is $\gamma \approx 74$ MHz, which corresponds to an average quality factor of $Q \approx 2.67 \times 10^6$. This value is significantly higher than the accuracy and resolution of the wavemeter used, thereby confirming the validity of the experimental protocol. From the 84 resonances fit (both TE and TM modes), we find a root-mean-square error per resonance center of $\sigma(\omega_s) = 1.90\,MHz$.
\begin{figure}[h]
    \centering
    \includegraphics[width=1\linewidth]{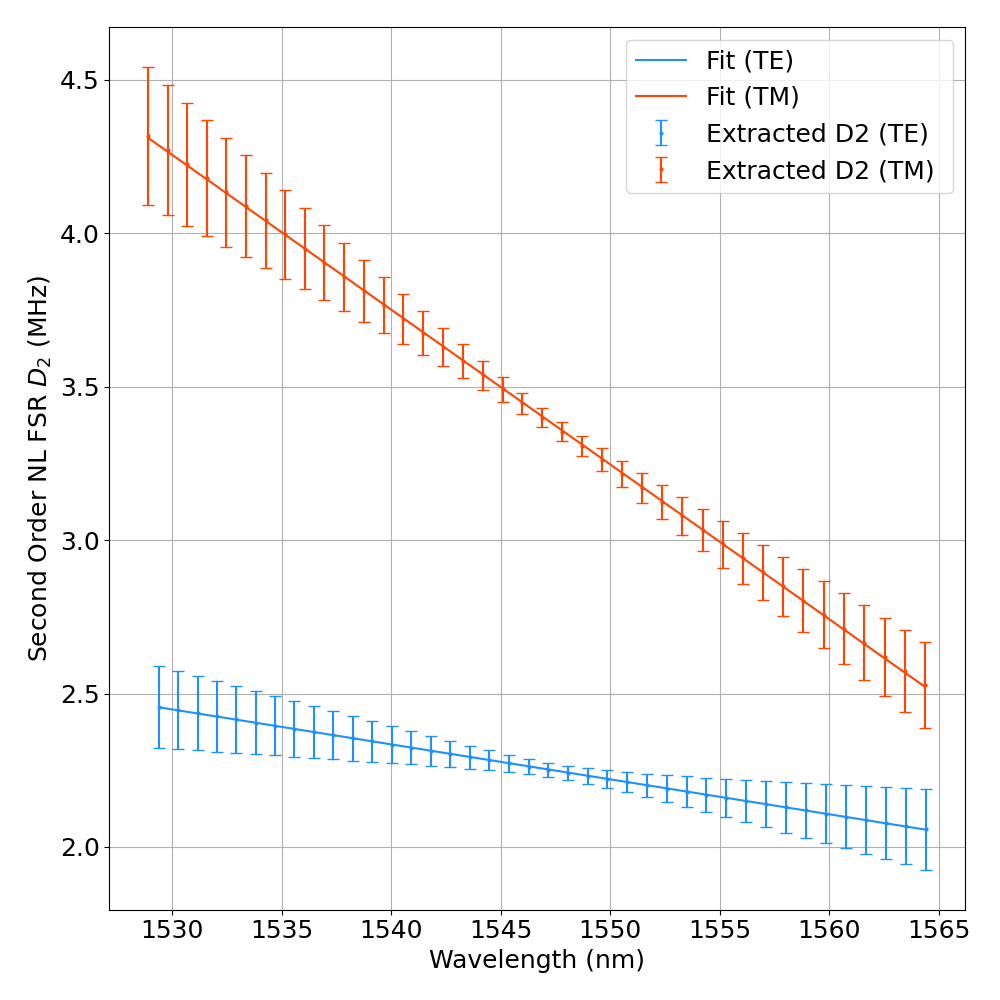}
    \caption{Extracted second order FSR $D_2$ for each scanned resonances for TE (blue) and TM (orange) modes. The lines are linear fits for extracting the dispersion slope $D_3$. The error bars correspond to $1\sigma$ uncertainty extracted from the fit using equation \ref{eq: dispersion microring}. }
    \label{fig: disp 2 et 3}
\end{figure}
Next, we use Eq.~\ref{eq: dispersion microring} to compute $D_{int}$ for both TE and TM modes for each scanned resonance,  as shown in Fig. ~\ref{fig: dint} (each serves as a reference for the Taylor expansion). Specifically, we extract the coefficient $D_2$ over the entire spectral range using Eq. ~\ref{eq: D2 microring}. The results are shown in Fig. \ref{fig: disp 2 et 3}. As expected, we observe a quasi-linear dependence of $D_2$ for both TE and TM modes over the wavelength range scanned by the probe laser. This behavior is consistent with the fact that most materials, including silicon nitride, exhibit a refractive index that varies slowly in the telecom band. Moreover, the MR under study was specifically designed to optimize the Kerr frequency comb generation at telecom wavelengths, which imposes the key conditions of anomalous dispersion \textit{i.e} a positive $D_{int}$, condition confirmed by our measurements. At the central resonance ($\lambda \approx 1547.169\,\text{nm}$), where the measurement uncertainty is minimal, we obtain for the TE mode a second order coefficient $D_2 = 2.24(2)\,\text{MHz}$, corresponding to a 1$\sigma$ uncertainty of about 1\%. To the best of our knowledge, this represents a state-of-the-art measurement of the FSR in a silicon nitride microresonator~\cite{eld19, Lucas2019Thesis, Chen2015Thesis}. By performing a linear fit of the FSR dependence, we extract the dispersion slope, yielding $D_3 \approx -3.4\,\text{kHz}$ and $-15.3\,\text{kHz}$ for the TE and TM modes, respectively. We emphasize that the main limitation of our measurements arises from the finite number of resonances accessible within the scanning range of the probe laser, which is directly related to the MR geometry. Increasing the number of accessible resonances would enhance the extraction of $D_{int}$, enabling a more precise determination of higher-order dispersion terms. We highlight that this method may be particularly relevant and of great interest for understanding the dynamics of frequency-comb generation in integrated structures~\cite{herr_universal_2012}. Moreover, it holds significant potential for engineering a flat dispersion profile, where dispersion variations must be maintained within tens of \(\text{ps}/\text{nm}/\text{km}\)~\cite{wang_ultra-flat_2016, Zhang:11, ChavezBoggio:14}.\\


Although the effective refractive index of the guided mode is not independently measured, the experimentally extracted values of \( D_2 \) and \( D_3 \) already constitute a direct and reliable experimental signature of the resonator’s spectral behavior. While the absolute dispersion values cannot be determined without precise knowledge of \( n_{\text{eff}}(\omega) \), the sign and spectral evolution of \( D_2 \) nonetheless provide clear insight into the dispersion regime and its wavelength dependence. Complementary numerical simulations could further strengthen this understanding by correlating the measured FSR coefficients with the waveguide geometry.  Accurately determining \( n_{\text{eff}}(\omega) \) remains challenging due to its strong sensitivity to nanometric geometric variations, yet this very sensitivity makes the dispersion analysis a powerful diagnostic tool. Indeed, the small variation of \( D_{2\omega} \), below \( 10\,\text{ps}/\text{nm}/\text{km} \) across the measured spectral range, is fully consistent with state-of-the-art fabrication tolerances~\cite{Zhang:11}. This demonstrates the excellent uniformity and control of the process, and conversely, such measurements can be exploited to reverse-engineer the waveguide geometry and finely calibrate the fabrication workflow. Finally, generating a Kerr frequency comb would provide a natural and experimentally robust benchmark, as its onset dynamics are directly governed by the chromatic dispersion parameters~\cite{herr_universal_2012}.

By leveraging the Taylor expansion of the resonance frequencies of a silicon nitride microresonator, we measured the FSR over a 5\,THz span, achieving a MHz resolution. Our approach, both robust and simple, relies solely on the precise measurement of resonance frequencies using a tunable laser and wavemeter. By considering both single- and double-peak resonances in our fitting process, we have achieved an accurate determination of the resonance centers. This level of precision enabled us to extract a highly resolved second-order FSR expansion parameter for both TE and TM modes across 42 resonances, from which we inferred the dispersion slope. Our findings establish a reliable method for characterizing MR dispersions, offering valuable insights for refining fabrication techniques and optimizing dispersion engineering in the development of on-chip frequency combs. Moreover, this technique could be extended to other material platforms or higher-order modes, providing a direct experimental tool for broadband dispersion engineering.


\vspace{20pt}

\section*{Acknowledgment}

This work was conducted within the framework of the OPTIMAL project granted by the European Union by means of the Fond Européen de développement régional (FEDER). The authors also acknowledge financial support from the Agence Nationale de la Recherche (ANR) through the SPHIFA project (ANR-20-CE47-0012), and the French government through its Investments for the Future programme under the Université Côte d'Azur UCA-JEDI project (Quantum@UCA) managed by the ANR (ANR-15-IDEX-01).\\

\section*{Competing interests}
The authors declare that they have no conflict of interest.

\section*{Data Availability}
The data are available from the authors upon request.

\end{document}